\documentclass[copyright]{eptcs}

\usepackage[utf8x]{inputenc}
\usepackage[T1]{fontenc}
\usepackage{amsmath}
\usepackage{amssymb}
\usepackage{theorem}
\usepackage[english]{babel}
\usepackage{gastex}
\usepackage{subfigure}
\usepackage{wasysym}
\usepackage{xspace}
\usepackage{multicol}

\renewcommand{\phi}{\varphi}
\renewcommand{\epsilon}{\varepsilon}

\newcommand{\safe}{\mathtt{safe}}
\newcommand{\Sound}{\mathtt{Sound}}
\newcommand{\Chance}{\mathtt{Chance}}

\newcommand{\bbN}{{\mathbb N}}
\newcommand{\bbP}{{\mathbb P}}

\newcommand{\calA}{{\cal A}}
\newcommand{\calD}{{\cal D}}
\newcommand{\calF}{{\cal F}}
\newcommand{\calG}{{\cal G}}
\newcommand{\calO}{{\cal O}}
\newcommand{\calP}{{\cal P}}
\newcommand{\calS}{{\cal S}}
\newcommand{\calV}{{\cal V}}
\newcommand{\calW}{{\cal W}}
\newcommand{\calX}{{\cal X}}
\newcommand{\calY}{{\cal Y}}

\newcommand{\tta}{{\tt a}}
\newcommand{\tti}{{\tt i}}
\newcommand{\ttu}{{\tt u}}

\newcommand{\game}{\calG}
\newcommand{\arena}{\calA}

\newcommand{\vertices}{\calV}
\newcommand{\actE}{\calX}
\newcommand{\actA}{\calY}
\newcommand{\trans}{\delta}
\newcommand{\colours}{\Gamma}
\newcommand{\col}{\gamma}
\newcommand{\sigE}{\Phi}
\newcommand{\obsE}{\phi}
\newcommand{\sigA}{\Psi}
\newcommand{\obsA}{\psi}

\newcommand{\play}{\pi}

\newcommand{\plays}{\Omega}
\newcommand{\os}{\plays^{\sigma}}
\newcommand{\ot}{\plays^{\tau}}
\newcommand{\oq}{\plays_{q}}
\newcommand{\ost}{\plays^{\sigma,\tau}}
\newcommand{\ostq}{\plays^{\sigma,\tau}_q}

\newcommand{\proba}[3]{\bbP^{#1,#2}_{#3}}
\newcommand{\pstq}{\proba{\sigma}{\tau}{q}}

\newcommand{\val}{\mathbf{v}}
\newcommand{\vs}{v_{\sigma}}
\newcommand{\vt}{v_{\tau}}
\newcommand{\vst}{v_{\sigma,\tau}}
\newcommand{\vstq}{\vst(q)}

\newcommand{\sigmaa}{\sigma^\tta}
\newcommand{\sigmau}{\sigma^\ttu}

\newcommand{\sigmai}{\sigma^\tti}

\newcommand{\pow}[1]{\calP(#1)}

\renewcommand{\iff}{if and only if\xspace}

\newcommand{\ie}{\textit{i.e.}\xspace}
\newcommand{\eg}{\textit{e.g.}\xspace}

\newenvironment{sketch}{{\noindent \it Sketch of proof.\/}}{\hfill $\square$ \medskip}

\newtheorem{theorem}{Theorem}

\newtheorem{definition}[theorem]{Definition}

\newcommand{\target}{\circledcirc}
\newcommand{\sink}{\otimes}

\title{How do we remember the past in randomised strategies?}
\author{Julien Cristau
\institute{LIAFA\\ CNRS \& Universit\'e Denis Diderot - Paris 7\\ Paris, France}
\email{jcristau@liafa.jussieu.fr}
\and Claire David 
\institute{\hphantom{CNRS \& Universit}LFCS\hphantom{is Diderot - Paris 7}\\ University of Edinburgh\\ Edinburgh, Scotland}
\email{cdavid@inf.ed.ac.uk}
\and Florian Horn
\institute{
LIAFA\\ CNRS \& Universit\'e Denis \\Diderot - Paris 7\\ Paris, France \and
IST Austria\\ Institute of Science and \\Technology Austria\\ Klosterneuburg, Austria
}
\email{horn@liafa.jussieu.fr}
}

\date\today

\begin{document}
\def\titlerunning{Remember the Past in Randomised Strategies}
\def\authorrunning{J. Cristau, C. David \& F. Horn}

\maketitle

\begin{abstract}
	Graph games of infinite length are a natural model for open reactive processes: one player represents the controller, trying to ensure a given specification, and the other represents a hostile environment. The evolution of the system depends on the decisions of both players, supplemented by chance.
	
	In this work, we focus on the notion of \emph{randomised} strategy. More specifically, we show that three natural definitions may lead to very different results: in the most general cases, an almost-surely winning situation may become almost-surely losing if the player is only allowed to use a weaker notion of strategy. In more reasonable settings, translations exist, but they require infinite memory, even in simple cases. Finally, some traditional problems becomes undecidable for the strongest type of strategies.
\end{abstract}

\bigskip

\bigskip

	\textit{``You can't have a strategy against telepaths: you have to act randomly. You have to not know what you're going to do next. You have to shut your eyes and run blindly. The problem is: how can you randomise your strategy, yet move purposefully towards your goal?''}

	\bigskip

	\hfill Solar Lottery

	\hfill Philip K. Dick \nocite{Dic55}

\section{Introduction}

	Since their introduction to verification in the late eighties, graph games have emerged as the model of choice for problems about \emph{open systems}, where a controller (Eve) must interact with an \emph{a priori} hostile environment (Adam)~\cite{PR89-POPL}. In such games, an \emph{arena} ---\ie a graph--- models the system and its evolution: at the beginning, a token is laid on one of the vertices, and its moves are determined by the actions of the players, supplemented by chance. The infinite sequence of vertices that ensues constitutes a \emph{play} of the game, whose winner is defined by some predetermined specification, often given as a regular condition on infinite words~\cite{MP92}.

	This model has been declined in a multiplicity of variants, in terms of both arenas and objectives. However, the questions are nearly always the same: Is there a winning strategy? For which player? How complex is it, in terms of memory and randomisation? Memory is the quantity of information that one is allowed to remember from the past: in general, the whole history is available, but it is often enough to remember a finite quantity of information. In addition, a strategy is \emph{pure} if it proposes only one possible action after any given sequence of observations. The notion of \emph{randomised} strategy, and its relation to memory, is the subject of this paper.

	 In verification, ``randomised strategy'' usually refers to a function from the history to a probability distribution over the actions. In other domains, such strategies are called ``\emph{behavioural} strategies'', as opposed to two other models of randomised strategies: a \emph{mixed} strategy is a measure over pure strategies, and a \emph{general} strategy is a measure over behavioural strategies. These models are also relevant in computer science. Indeed, the IPv6 ``Stateless Address Autoconfiguration'' protocol, which only uses randomisation at the beginning to generate a new I.P. address~\cite{TNJ07-RFC}, can be accurately described as a mixed strategy. Likewise, the secure shell protocol (ssh) is a general strategy, since a new session key has to be randomly\footnote{except on Debian~\cite{BB08-DEFCON}.} generated every hour or gigabyte~\cite{YL06-RFC}.

	 In this paper, we propose definitions for mixed and general strategies, with or without memory, in the framework of graph games for verification. We expose several situations in which their analyses differ significantly from the behavioural model. In the most general case, the same game can be almost-surely losing or almost-surely winning depending on the type of strategies we consider. In other situations, we conjecture that the values are the same, but we show that memory needs vary (from two to infinity). Altogether, we hope to ask more questions than we give answers: our main objective is to describe these three models for randomised strategies in graph games and to point out that many problems which are solved for behavioural strategies are still open in the mixed and general cases.

	 The paper is organised as follows. In Section~\ref{section:definitions}, we recall the classical notions about graph games in verification, in a very general framework which subsumes a large part of the literature. Section~\ref{section:randomisedstrategies} presents our definitions for behavioural, mixed, and general strategies in graph games, and stresses the fundamental differences between the three notions. Section~\ref{section:memory} focuses on memory-related issues: it exhibits variations in the elementary cases of concurrent safety games and simple Muller games. In Section~\ref{section:discussion}, we sum up our observations and results, and propose some open problems.

\section{Definitions}
\label{section:definitions}

	\subsubsection*{Notation}

	For a finite or countable set $\calS$, we denote by $\calD(\calS)$ the set of probability distributions over $\calS$, \ie the set of functions from $\calS$ to positive real numbers that sums up to one.

	\subsubsection*{Arenas and plays}

	An \emph{arena} $\arena$ is a tuple $(\vertices, \actE, \actA, \trans, \colours, \col, \sigE, \obsE, \sigA, \obsA)$ where $\vertices$ is the set of vertices in the graph, $\actE$ is the set of actions of Eve, $\actA$ is the set of actions of Adam, $\trans : \vertices \times \actE \times \actA \rightarrow \calD(\vertices)$ is the transition function, $\colours$ is the set of colours, $\col: \vertices \rightharpoonup \colours$ is the colouring function, $\sigE$ is the set of signals of Eve, $\obsE: \vertices \cup \actE \rightharpoonup \sigE$ is her observation function, $\sigA$ is the set of signals of Adam, and $\obsA: \vertices \cup \actA \rightharpoonup \sigA$ is his observation function. Many results about graph games for verification consider only restricted arenas, such as:
	\begin{description}
		\item{\bf Synchronous:} an arena is \emph{synchronous} if $\obsE$ and $\obsA$ are total.
		\item{\bf Observable actions:} a synchronous arena has \emph{observable actions} if the restrictions of $\obsE$ to $\actE$ and $\obsA$ to $\actA$ are one-to-one.
		\item{\bf Perfect information:} a synchronous arena has \emph{perfect information} (or is \emph{concurrent}) if the restrictions of $\obsE$ and $\obsA$ to $\vertices$ are one-to-one.
		\item{\bf Simple:} a concurrent arena is \emph{simple} (or \emph{turn-based}) if for each vertex $q$, $\trans(q,x,y)$ depends either on $x$ or on $y$, but not both.
	\end{description}
	A \emph{play} on the arena $\arena$ is a (possibly infinite) sequence $\play = \play_0 \play_1 \ldots$ of states such that $\forall i < |\play|\! -\! 1,\exists x \in \actE, y \in \actA, \trans(\play_i,x,y)(\play_{i+1}) > 0$. The set of plays is usually denoted $\plays$, and the set of plays starting with the vertex $q$ by $\oq$.

	\subsubsection*{Pure strategies and measures} A \emph{pure strategy $\sigma$} for Eve (resp. $\tau$ for Adam) on the arena $\arena$ associates an action to each finite sequence of observations: $\sigma : \sigE^* \rightarrow \actE$ (resp. $\tau : \sigA^* \rightarrow \actA$) . A play $\play$ is \emph{consistent with a strategy $\sigma$ for Eve} (resp. \emph{$\tau$ for Adam}) \iff at each step $i$, there is an action $y \in \actA$ for Adam (resp. $x \in \actE$) such that $\trans(\play_i,\sigma(\obsE(\play_{1..i})),y)(\play_{i+1}) > 0$ (resp. $\trans(\play_i,x,\tau(\obsA(\play_{1..i})))(\play_{i+1}) > 0$). Notice that, in the case of an asynchronous arena, actions can only change with new observations: otherwise, the same argument leads to the same result over and over. The set of plays consistent with $\sigma$ (resp. $\tau$; $\sigma$ and $\tau$) is denoted by $\os$ (resp. $\ot$; $\ost$). Once an initial vertex $q$ and two strategies $\sigma$ and $\tau$ have been fixed, $\ostq$ can naturally be made into a measurable space $(\ostq,\calO)$, where $\calO$ is the $\sigma$-field generated by the cones $\{\calO_w \mid w \in \vertices^*\}$: $\play \in \calO_w$ \iff $w$ is a prefix of $\play$. The probability measure $\pstq$ is recursively defined by $\pstq(\calO_q) = 1$ and:
		  \begin{displaymath}
			\forall w \in \vertices^*, (r,s) \in \vertices^2, \pstq(\calO_{wrs}) = \pstq(\calO_{wr}) \cdot \trans(r,\sigma(\obsE(wr)),\tau(\obsA(wr)))(s) .
		  \end{displaymath}
	Carath\'eodory's extension theorem allows us to extend $\pstq$ to the Borel sets of $(\ostq,\calO)$~\cite{Wil91}.

	\subsubsection*{Winning conditions and values} A \emph{winning condition $\calW$ on a set of colours $\colours$} is a Borel subset of $\colours^\infty$. A play $\play$ in an arena $\arena$ on $\colours$ is \emph{winning for Eve} in the game $(\arena,\colours)$ if $\col(\play) \in \calW$, and \emph{winning for Adam} otherwise. In a game $\game = (\arena,\calW)$, the \emph{pure value of a state $q$ under the strategies $\sigma$ and $\tau$}, denoted $\vst(q)$, is the measure of $\calW$ under $\pstq$. The \emph{value for Eve of a state $q$} is the supremum of the values that she can ensure from $q$ against any strategy of Adam. Symmetrically, the \emph{value for Adam of a state $q$} is the infimum of the values he can defend against any strategy of Eve. In simple stochastic games, these two values coincide and are usually called the \emph{value of $q$}~\cite{Mar98-JSL,MS98-IJGT}\footnote{As a matter of fact, these papers shows the quantitative determinacy, in behavioural strategies, of Borel games on concurrent arenas. An inspection of the proof yields the same result for pure strategies in the case of simple arenas.}. 
		\begin{displaymath}
			\val(q) = \sup_\sigma \vs(q) = \inf_\tau \vt(q) \enspace .
		\end{displaymath}

	\subsubsection*{Winning criteria} Following de Alfaro and Henzinger~\cite{dAH00-LICS}, we consider several notions of \emph{winning strategies} and \emph{winning regions}, depending on the chances Eve has to win. In decreasing order of difficulty, and from an initial vertex $q$, a strategy $\sigma$ for Eve:
	\begin{itemize}
		\item is \emph{sure} if any play consistent with $\sigma$ is winning for Eve;
		\item is \emph{almost-sure} if for any strategy $\tau$ for Adam, $\vstq = 1$;
		\item \emph{ensures $\epsilon$} if for any strategy $\tau$ for Adam, $\vstq \ge \epsilon$;
		\item is \emph{positive} if for any strategy $\tau$ for Adam, $\vstq > 0$;
		\item is \emph{heroic} if for any strategy $\tau$, there is a play consistent with $\sigma$ and $\tau$ which is winning for Eve. 
	\end{itemize}
	The \emph{sure region} (resp. \emph{almost-sure region of Eve}, \emph{positive region}, \emph{heroic region}) of Eve is the set of vertices from which she has a sure (resp. almost-sure, positive, heroic) strategy. Furthermore, the \emph{bounded region} is the set of vertices from which Eve has a strategy ensuring a positive $\epsilon$ and the \emph{limit-one region} is the set of vertices from which Eve has a strategy ensuring $\epsilon$ for any $\epsilon < 1$. The same concepts are defined accordingly for Adam, except that we say that a strategy $\tau$ for Adam \emph{defends $\epsilon$} if it guarantees that, for any strategy $\sigma$ for Eve, $\vstq \le \epsilon$.

\section{Behavioural, mixed, and general strategies}
\label{section:randomisedstrategies}

	As soon as we deal with concurrent arenas, we cannot rely only on pure strategies to make meaningful analyses. In the classical game of ``Janken'', any pure strategy is surely beaten by the appropriate counter-strategy ($\mathtt{paper}$ against $\mathtt{rock}$, $\mathtt{scissors}$ against $\mathtt{paper}$, and $\mathtt{rock}$ against $\mathtt{paper}$), but a strategy which plays each action with probability $\frac{1}{3}$ eventually wins with probability $\frac{1}{2}$. The main point of this paper is that there are several possible definitions for the notion of ``randomized strategy''.
	\begin{itemize}
		\item	A \emph{behavioural strategy} returns at each step a distribution over the actions: $\sigE^* \rightarrow \calD(\actE)$;
		\item	a \emph{mixed strategy} is a measure over pure strategies: $\calD(\sigE^* \rightarrow \actE)$;
		\item	a \emph{general strategy} is a measure over behavioural strategies: $\calD(\sigE^* \rightarrow \calD(\actE))$.
	\end{itemize}
	As we show in this paper the expressive powers of these models are quite different. Intuitively, a behavioural strategy does not know in advance what it will play next, so its actions can change when its decisions do not (even when there are no observations). Mixed strategies use randomization to get hidden information at the beginning of the play, which can later be used to correlate undistinguishable actions, \eg playing $aa$ or $bb$ with probability $\frac{1}{2}$. General strategies subsume both, so they can, in particular, generate hidden information on the fly. These distinctions have mostly been overlooked in verification (apart from a few remarks, \eg \cite{dAHK98-FOCS,DHR08-IJFCS}). One reason is that the games we consider are usually synchronous, with observable actions. On synchronous arenas, mixed strategies can simulate behavioural strategies: as each action can be uniquely identified beforehand by its position in the play, it is possible to define a measure which somehow makes all the random draws at the beginning of the play. If furthermore, the actions are observable, Kuhn's theorem states that mixed and behavioural (and thus general) strategies have the same expressive power~\cite{Aum64-AMST}.

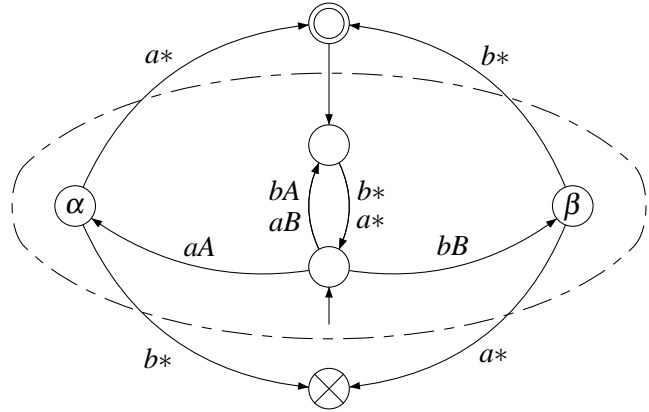
\begin{figure}[ht]
		\begin{center}
		\unitlength=1.35mm
			\begin{picture}(70,42)(0,0)
				\gasset{Nh=4,Nw=4,polyangle=90}
				\node[Nmarks=i, iangle=270](i)(35,15){}
				\node(delay)(35,27){}
				\node(a)(10,21){$\alpha$}
				\node(b)(59,21){$\beta$}
				\node[Nmarks=r](top)(35,39){}
				\node(bottom)(35,3){}
				\drawline[AHnb=0](36.45,4.45)(33.55,1.55){}
				\drawline[AHnb=0](36.45,1.55)(33.55,4.45){}
				\drawedge[curvedepth=2,ELpos=40](i,delay){$aB$}
				\drawedge[curvedepth=2,ELpos=60](i,delay){$bA$}
				\drawedge[curvedepth=3,ELside=r](i,a){$aA$}
				\drawedge[curvedepth=-3](i,b){$bB$}
				\drawedge[curvedepth=2,ELpos=40](delay,i){$b*$}
				\drawedge[curvedepth=2,ELpos=60](delay,i){$a*$}
				\drawedge[curvedepth=5](a,top){$a*$}
				\drawedge[curvedepth=-5,ELside=r](a,bottom){$b*$}
				\drawedge[curvedepth=-5,ELside=r](b,top){$b*$}
				\drawedge[curvedepth=5](b,bottom){$a*$}
				\drawedge(top,delay){}
				\drawccurve[dash={3 1 1 1}0](5,25)(5,17)(65,17)(65,25)
			\end{picture}
		\end{center}
		\caption{Who wins?}
		\label{figure:valuefails}
	\end{figure}

	These hypotheses have been inconspicuously challenged in recent papers. In this regard, the comparison between~\cite{BGG09-LICS} and~\cite{GS09-ICALP} makes for an enlightening example. At first glance, these two papers look very similar: they both ponder the problem of the existence of almost-sure strategies in games where both players have (asymmetric) imperfect information. A closer examination reveals the differences: Bertrand, Genest, and Gimbert use general strategies, while Gripon and Serre use behavioural strategies; furthermore, in the latter paper, the players cannot observe their own actions. As a consequence, there are cases where the answer to the synthesis problem depends on which model is used. Consider for example the synchronous arena depicted in Figure~\ref{figure:valuefails}, where Eve cannot distinguish vertices nor actions in the dashed area, $\sink$ is a losing sink state, and $\circledcirc$ is her ``target'', for either a reachability or a B\"uchi condition.

	With a behavioural strategy, Eve's strategy can only depend on the length of the play. At any even move, if her strategy is to play $a$ with probability $p$ and $b$ with probability $1-p$, Adam can answer by playing $A$ with probability $1-p$ and $B$ with probability $p$, so the odds of the token going to $\alpha$ or to $\beta$ are equal (they are worth $p \cdot (1-p)$ each). In the next step, no matter what $\sigma$ advocates, the odds of the token going to $\target$ or to $\sink$ will again be equal. In the reachability game, this limits Eve's prospects to half chances. In the B\"uchi game, the probability that she wins drops to $0$.
	
	On the flip side, she has an almost-sure mixed strategy for both objectives: the natural ``uniform'' measure over the strategies of the form $(aa|bb)^\omega$ guarantees that each sequence of two moves starting in the initial vertex has a probability of $\frac{1}{2}$ to send the token to $\target$, and a probability of $\frac{1}{2}$ to send the token back to the initial vertex, no matter what Adam does. It cannot go to $\sink$, as Eve never plays $ab$ or $ba$ from the initial vertex.

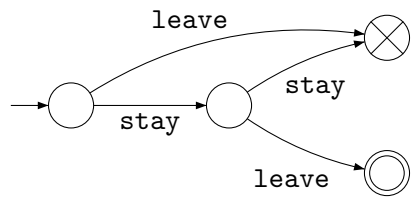
\begin{figure}[ht]
		\begin{center}
		\unitlength=1.5mm
			\begin{picture}(42,18)(0,0)
				\gasset{Nh=4,Nw=4,polyangle=90}
				\node[Nmarks=i, iangle=180](first)(7,9){}
				\node(second)(21,9){}
				\node[Nmarks=r](exit)(35,3){}
				\node(fail)(35,15){}
				\drawline[AHnb=0](36.45,16.45)(33.55,13.55){}
				\drawline[AHnb=0](36.45,13.55)(33.55,16.45){}
				\drawedge[ELside=r](first,second){$\mathtt{stay}$}
				\drawedge[curvedepth=1,ELside=r](second,fail){$\mathtt{stay}$}
				\drawedge[ELpos=42,curvedepth=3](first,fail){$\mathtt{leave}$}
				\drawedge[curvedepth=-1,ELside=r](second,exit){$\mathtt{leave}$}
			\end{picture}
		\end{center}
		\caption{The D.U.I. game}
		\label{figure:drinkanddrive}
	\end{figure}

	The arena of Figure~\ref{figure:valuefails} is synchronous, so any behavioural strategy can be emulated by a mixed one. If we remove this hypothesis, it is not always the case, as in the one-player game of Figure~\ref{figure:drinkanddrive}, where Eve is unaware of any action or vertex.

	As Eve observes nothing, her strategy is completely determined by what she does on the empty word $\lambda$. She has only two pure strategies: $\lambda \rightarrow \mathtt{stay}$ and $\lambda \rightarrow \mathtt{leave}$. Both lead to $\sink$, and so does any mixed strategy of the form $\{p \cdot (\lambda \rightarrow \mathtt{stay}), (1-p) \cdot (\lambda \rightarrow \mathtt{leave})\}$. The behavioural strategy $\lambda \rightarrow \{\frac{1}{2} \cdot \mathtt{stay}, \frac{1}{2} \cdot \mathtt{leave}\}$, on the other hand, yields one chance out of four to to reach $\target$.

	The case of games with perfect information and invisible actions is still open: there are mixed strategies which cannot be imitated by any behavioural one, so we cannot hope for a ``generic'' translation. But that does not rule out the possibility of specific, objective-dependent constructions which would yield a different strategy with the same value.

\section{Memory issues}
\label{section:memory}

	A refinement of the synthesis problem asks that the controller uses only finite memory, as a natural requirement for implementability. Pure strategies with memory are defined in the following way: 
	\begin{definition}
	\label{definition:purememory}
		A \emph{pure strategy $\sigma$ with memory $M$} is a triple $(\sigmai, \sigmau, \sigmaa)$ where $\sigmai \in M$ is the initial memory state; $\sigmau : (\sigE \times M) \rightarrow M$ is the \emph{memory update} function, which maps a signal and a memory state to a new memory state and is called at each new observation of Eve; and $\sigmaa : (\sigE \times M) \rightarrow \actE$ is the \emph{next-action} function, which maps a signal and a memory state to an action and is called at each step.
	\end{definition}
	Notice that any pure strategy $\sigma$ can be represented as a strategy with memory $\sigE^*$, with $\sigmai = \lambda$, $\sigmau = \cdot$ and $\sigmaa = \sigma$. A strategy has \emph{finite memory} if $M$ is a finite set, and is \emph{memoryless} if $M$ is a singleton.

	Randomized strategies with (countable) memory are defined with similar tuples, except that some of their elements use randomization.
	\begin{description}
		\item{\bf Behavioural:}	In a \emph{behavioural strategy with memory $M$}, the next-action $\sigmaa : (\sigE \times M) \rightarrow \calD(\actE)$ is randomized.
		\item{\bf Mixed:}	In a \emph{mixed strategy with memory $M$}, the initial memory $\sigmai \in \calD(M)$ is randomised.
		\item{\bf General:}	In a \emph{general strategy with memory $M$}, the next-action, initial memory, and memory-update $\sigmau : (\sigE \times M) \rightarrow \calD(M)$ are randomised.
	\end{description}

	The memory requirements can also depend of the type of strategy. In the game of Figure~\ref{figure:valuefails}, for example, there is no almost-sure mixed strategy with finite memory (in the reachability game, there are $\epsilon$-optimal strategies with finite (unbounded) memory; in the B\"uchi game, every mixed strategy with finite memory has value $0$). However, the strategy we described can be realized by a general strategy with four memory states $\mathtt{a_{even}}$, $\mathtt{a_{odd}}$, $\mathtt{b_{even}}$ and $\mathtt{b_{odd}}$: in the $\mathtt{even}$ memory state, she updates her memory at random to one of the $\mathtt{odd}$ states; in the $\mathtt{odd}$ states, she updates her memory to the corresponding $\mathtt{even}$ state; in all states, she plays the action corresponding to her memory state.

	\subsection{Concurrent safety games}

\begin{figure}[ht]
		\begin{center}
		\unitlength = 3.50mm
			\begin{picture}(24,8)(0,0)
				\gasset{Nw=2,Nh=2,Nmr=6,loopdiam=2.8}
				\node(un)(2,3){}
				\node[Nmarks=i,iangle=270](deux)(12,3){}
				\node(trois)(22,3){}
				\drawline[AHnb=0](22.7,3.7)(21.3,2.3){}
				\drawline[AHnb=0](21.3,3.7)(22.7,2.3){}				
				\drawedge[ELside=r](deux,un){$\mathtt{throw}|\mathtt{run}$}
				\drawloop(deux){$\mathtt{wait}|\mathtt{hide}$}
				\drawedge(deux,trois){$\mathtt{wait}|\mathtt{run}$}
				\drawedge[ELside=r](deux,trois){$\mathtt{throw}|\mathtt{hide}$}
			\end{picture}
		\end{center}
		\caption{The snowball game}
		\label{figure:snowball}
	\end{figure}
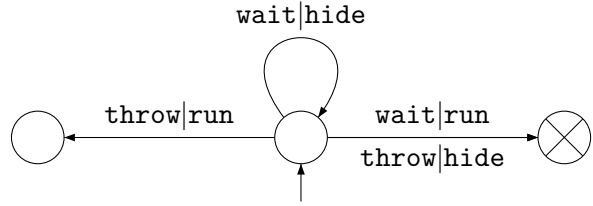

	In~\cite{dAHK98-FOCS}, de~Alfaro, Henzinger, and Kupferman study the problem of concurrent reachability/safety games and establish the qualitative determinacy of these games, as well as several results on the nature (memory and randomization) of the strategies needed to achieve various objectives. In particular, they show that positive strategies for safety objectives require, in general, an infinite amount of memory. The proof is based on the famous ``snowball game'' of~\cite{KS81-SJCO}, which is pictured in Figure~\ref{figure:snowball}.

	In this game, Adam loses if he never runs and Eve never throws, or if Eve happens to throw the snowball exactly at the moment he runs. It is clear that Adam has memoryless behavioural strategies with value arbitrarily close to one: if, at each step, he chooses to run with probability $\epsilon$, he ensures a probability of winning of $1-\epsilon$ (Eve's best chance is to throw the ball right away). It is also clear that he cannot win almost-surely: if he has a positive probability of never running, Eve can keep the snowball forever; and if he has a positive probability of running at any step, Eve can thwart him by throwing the ball with probability $\frac{1}{2}$ at each step.

	By the \emph{qualitative determinacy} of concurrent regular games, Eve has a positive strategy, \ie a unique strategy which prevents Adam from winning almost-surely with any strategy. De~Alfaro, Henzinger, and Kupferman use behavioural strategies, and argue that Eve needs infinite memory: the sequence $(\sigma(\Circle^i)(\mathtt{throw}))_{i\in\bbN}$ must go to $0$ but never reach it. It is clear that there are no positive mixed strategies with finite memory, as pure strategies with $n$ memory states can only throw the snowball in the first $n$ steps.

	On the other hand, there is a general strategy with only $2$ memory states: in the memory state $\mathtt{Never}$, Eve keeps the snowball with probability $1$, in the state $\mathtt{Eventually}$, she throws it with probability $\frac{1}{2}$; the memory never changes, and the initial memory state is chosen at random. This strategy prevents Adam from winning almost-surely, since he can never be sure that Eve is not in the memory state $\mathtt{Eventually}$. In fact, this is the case in every finite concurrent safety game:

	\begin{theorem}
	\label{theorem:concurrentsafetymixed}
		In every finite concurrent safety game, Eve has a positive general strategy with memory $2$ from her positive region.
	\end{theorem}

	\begin{sketch}
		It follows from the analysis of the fix-points in~\cite{dAHK98-FOCS} that there is a total preorder~$\prec$ on the vertices such that:
		\begin{itemize}
			\item	the minimal vertices belong to the almost-sure region of Adam;
			\item	for each non-minimal vertex $q \in \vertices$, there is an action $\safe_q \in \actE$ of Eve such that, for any action $y$ of Adam:
				\begin{itemize}
					\item	either for any vertex $r \in \vertices$, $\trans(q,\safe_q,y)(r) > 0 \Rightarrow q \precsim r$,
					\item	or there is an action $x \in \actE$ of Eve and a vertex $r \in \vertices$ such that $\trans(q,x,y)(r) > 0 \wedge q \precnsim r$
				\end{itemize}
		\end{itemize}
		Notice that always playing the $\safe$ action is a pure and positional sure strategy for Eve in the maximal vertices (unless they are also minimal). For the vertices in between, we claim that the following strategy with two memory states is positive for Eve:
		\begin{itemize}
			\item	the memory states are called $\Sound$ and $\Chance$;
			\item	each time the token goes to a new $\prec$-class, the memory state is updated to either $\Sound$ or $\Chance$ with equal probabilities; otherwise, the memory does not change;
			\item	in the $\Sound$ state, Eve always plays the $\safe$ action of the current vertex;
			\item	in the $\Chance$ state, Eve plays any action in $\actE$ with equal probabilities.
		\end{itemize}
		The situation is roughly the same as in the snowball game: if Adam's actions have no chance to go to a lower vertex against the $\Sound$ strategy, he will lose with probability $\frac{1}{2}$; if he takes a risk at any point, there is a positive probability that Eve was in the $\Chance$ memory state all along, so he could end up in a greater vertex.
	\end{sketch}

	In addition to the finite memory, the strategy described in the proof of Theorem~\ref{theorem:concurrentsafetymixed} is simple, generic, and uses only uniform probabilities. By comparison, the description of a positive behavioural strategy is in general very complex and uses probabilities of unbounded precision.

	\subsection{Memory bounds for Muller games}

	Even in the elementary case of simple Muller games, it is not clear that the memory needs are the same for behavioural and general strategies. Recall that a simple arena is an arena with turn-based moves and perfect information for both players, and a Muller condition is a condition depending only on the set of colours visited infinitely often:

	\begin{definition}
		A Muller condition on a set of colours $\colours$ is specified by a subset $\calF$ of $\pow{\colours}$. A play $\play$ satisfies $\mathtt{Muller}(\calF)$ \iff the set of colours occurring infinitely often in $\col(\play)$ belongs to $\calF$.
	\end{definition}
	
	In such games, both players have pure optimal strategies with finite memory~\cite{BL69-TAMS}. A follow-up problem is to determine, for a given Muller condition $\calF$ on a set of colours $\colours$, the necessary and sufficient amount of memory needed to define optimal pure strategies in \emph{any} arena coloured by $\colours$. Gurevich and Harrington used the \textit{latest appearance record} (LAR) structure of McNaughton to give a first upper bound of $|\colours|!$~\cite{GH82-STOC}. Zielonka refined the LAR into a tree, whose leaves could be used as memory~\cite{Zie98-TCS}. Finally, Dziembowski, Jurdzinski, and Walukiewicz showed that each player needs only as much memory as the number of leaves in some particular sub-trees, establishing tight and asymmetrical bounds for pure strategies~\cite{DJW97-LICS}.

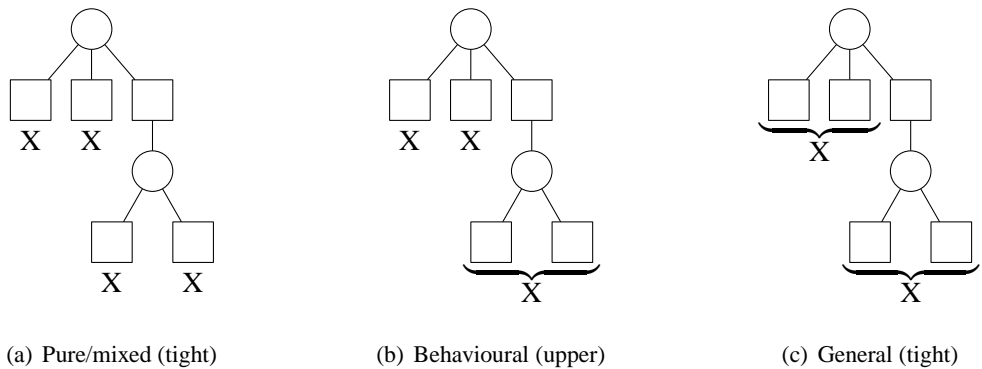
\begin{figure}[ht]
		\begin{center}
		\unitlength = 1.35mm
			\subfigure[Pure/mixed (tight)]{
			\label{subfigure:pure}
			\begin{picture}(28,29)(0,0)
	
				\gasset{Nw=4,Nh=4,AHnb=0}
	
				\node[Nmr=4](abcd)(12,27){}
				\node[Nmr=0](bcd)(6,20){}
				\node[Nmr=0](acd)(12,20){}
				\node[Nmr=0](abd)(18,20){}
				\node[Nmr=3](ab)(18,13){}
				\node[Nmr=0](a)(14,6){}
				\node[Nmr=0](b)(22,6){}
	
				\drawedge(abcd,bcd){}
				\drawedge(abcd,acd){}
				\drawedge(abcd,abd){}
				\drawedge(abd,ab){}
				\drawedge(ab,b){}
				\drawedge(ab,a){}
	
				\node[Nframe=n](1)(6,16){X}
				\node[Nframe=n](2)(12,16){X}
				\node[Nframe=n](3)(14,2){X}
				\node[Nframe=n](4)(22,2){X}
			\end{picture}
			} \hfil \subfigure[Behavioural (upper)]{
			\label{subfigure:behavioural}
			\begin{picture}(28,29)(0,0)
	
				\gasset{Nw=4,Nh=4,AHnb=0}
	
				\node[Nmr=4](abcd)(12,27){}
				\node[Nmr=0](bcd)(6,20){}
				\node[Nmr=0](acd)(12,20){}
				\node[Nmr=0](abd)(18,20){}
				\node[Nmr=3](ab)(18,13){}
				\node[Nmr=0](a)(14,6){}
				\node[Nmr=0](b)(22,6){}
	
				\drawedge(abcd,bcd){}
				\drawedge(abcd,acd){}
				\drawedge(abcd,abd){}
				\drawedge(abd,ab){}
				\drawedge(ab,b){}
				\drawedge(ab,a){}
	
				\node[Nframe=n](1)(6,16){X}
				\node[Nframe=n](2)(12,16){X}
				\node[Nframe=n](3)(18,3){$\underbrace{\hspace{1.8cm}}$}
				\node[Nframe=n](4)(18,1){X}
			\end{picture}
			} \hfil \subfigure[General (tight)]{
			\label{subfigure:mixed}
			\begin{picture}(28,29)(0,0)
	
				\gasset{Nw=4,Nh=4,AHnb=0}
	
				\node[Nmr=4](abcd)(12,27){}
				\node[Nmr=0](bcd)(6,20){}
				\node[Nmr=0](acd)(12,20){}
				\node[Nmr=0](abd)(18,20){}
				\node[Nmr=3](ab)(18,13){}
				\node[Nmr=0](a)(14,6){}
				\node[Nmr=0](b)(22,6){}
	
				\drawedge(abcd,bcd){}
				\drawedge(abcd,acd){}
				\drawedge(abcd,abd){}
				\drawedge(abd,ab){}
				\drawedge(ab,b){}
				\drawedge(ab,a){}
	
				\node[Nframe=n](1)(9,17){$\underbrace{\hspace{1.6cm}}$}
				\node[Nframe=n](2)(9,15){X}
				\node[Nframe=n](3)(18,3){$\underbrace{\hspace{1.8cm}}$}
				\node[Nframe=n](4)(18,1){X}
			\end{picture}
			}
	
		\end{center}
	\caption{Memory bounds for simple Muller games}
	\label{figure:mullerbounds}
	\end{figure}

	It is clear from their proof that mixing strategies does not help, since the other player can efficiently adapt their strategy in the witness arenas. This is not the case for behavioural strategies: Chatterjee, de Alfaro, and Henzinger observed that upward-closed winning conditions admitted memoryless strategies~\cite{CdAH04-QEST}, leading to smaller upper bounds for arbitrary Muller conditions~\cite{Cha07-FoSSaCS}. Horn established even smaller tight bounds for general strategies~\cite{Hor09-STACS} (see Figure~\ref{figure:mullerbounds} for a graphical representation of the three bounds on a Zielonka (sub-)tree). However, Horn's upper-bound has not yet been proven (or refuted) for behavioural strategies.

\section{Discussion}
\label{section:discussion}
	We have compared three models of randomized strategies ---behavioural, mixed, and general--- in the context of graph games. Depending on the sub-case, we were able to expose variations in the amount of memory needed, the existence of finite-memory strategies, or even the values. In concurrent games with unobservable actions, the equivalence between the three models is still an open question.
	
	In verification, the behavioural model has received most of the attention. Nevertheless, there is \emph{a priori} nothing wrong with the other types of controllers. Furthermore, in several cases, general strategies can be much simpler than behavioural or mixed ones. On the other hand, general strategies are much less amenable to further analysis, as they introduce imperfect information. Even in simple safety games, one cannot compute the value of a general strategy ---or even decide if it has positive value~\cite{GO09-LaBRI}.
	
	Each model has strengths and weaknesses, and we do not favour one over the others. Our point is rather to stress the importance of this initial choice, and to note that many memory-related problems which have been solved for behavioural strategies are still open in the mixed and general frameworks.

\end{document}